\documentclass[twoside]{article}
\usepackage{epsfig}

\catcode`\@=11
\long\def\@makefntext#1{
\protect\noindent \hbox to 3.2pt {\hskip-.9pt
$^{{\eightrm\@thefnmark}}$\hfil}#1\hfill}		

\def\thefootnote{\fnsymbol{footnote}}
\def\@makefnmark{\hbox to 0pt{$^{\@thefnmark}$\hss}}	
	
\def\ps@myheadings{\let\@mkboth\@gobbletwo
\def\@oddhead{\hbox{}
\rightmark\hfil\eightrm\thepage}
\def\@oddfoot{}\def\@evenhead{\eightrm\thepage\hfil
\leftmark\hbox{}}\def\@evenfoot{}
\def\sectionmark##1{}\def\subsectionmark##1{}}

\oddsidemargin=\evensidemargin
\renewcommand{\thefootnote}{\fnsymbol{footnote}}
\newcommand{\alphfootnote}
	{\setcounter{footnote}{0}
	 \renewcommand{\thefootnote}{\sevenrm\alph{footnote}}}

\usepackage{titlesec}
\titleformat{\section}{\normalsize \bfseries}{\thesection}{1em}{}
\titleformat{\subsection}{\small \it \bfseries}{\thesubsection}{1em}{}

\newcommand{\nonumsection}[1] {\vspace{12pt}\noindent{\tenbf #1}
	\par\vspace{5pt}}

\newcounter{appendixc}
\newcounter{subappendixc}[appendixc]
\newcounter{subsubappendixc}[subappendixc]
\renewcommand{\thesubappendixc}{\Alph{appendixc}.\arabic{subappendixc}}
\renewcommand{\thesubsubappendixc}
	{\Alph{appendixc}.\arabic{subappendixc}.\arabic{subsubappendixc}}

\renewcommand{\appendix}[1] {\vspace{12pt}
        \refstepcounter{appendixc}
        \setcounter{figure}{0}
        \setcounter{table}{0}
        \setcounter{lemma}{0}
        \setcounter{theorem}{0}
        \setcounter{corollary}{0}
        \setcounter{definition}{0}
        \setcounter{equation}{0}
        \renewcommand{\thefigure}{\Alph{appendixc}.\arabic{figure}}
        \renewcommand{\thetable}{\Alph{appendixc}.\arabic{table}}
        \renewcommand{\theappendixc}{\Alph{appendixc}}
        \renewcommand{\thelemma}{\Alph{appendixc}.\arabic{lemma}}
        \renewcommand{\thetheorem}{\Alph{appendixc}.\arabic{theorem}}
        \renewcommand{\thedefinition}{\Alph{appendixc}.\arabic{definition}}
        \renewcommand{\thecorollary}{\Alph{appendixc}.\arabic{corollary}}
        \renewcommand{\theequation}{\Alph{appendixc}.\arabic{equation}}
        \noindent{\tenbf Appendix \theappendixc #1}\par\vspace{5pt}}
\newcommand{\subappendix}[1] {\vspace{12pt}
        \refstepcounter{subappendixc}
        \noindent{\bf Appendix \thesubappendixc. {\kern1pt \bfit #1}}
	\par\vspace{5pt}}
\newcommand{\subsubappendix}[1] {\vspace{12pt}
        \refstepcounter{subsubappendixc}
        \noindent{\rm Appendix \thesubsubappendixc. {\kern1pt \tenit #1}}
	\par\vspace{5pt}}

\newcommand{\textlineskip}{\baselineskip=13pt}
\newcommand{\smalllineskip}{\baselineskip=10pt}

\newcommand{\copyrightheading}[1]
	{\vspace*{-2.5cm}\smalllineskip{\flushleft
	{\footnotesize Quantum Information and Computation, Vol.~1, No.~0 (2001) 000--000 #1}\\
	{\footnotesize \copyright\kern2pt Rinton Press}\\
	 }}

\def\abstracts#1#2#3{{
	\centering{\begin{minipage}{4.5in}\footnotesize\baselineskip=10pt
	\parindent=0pt #1\par
	\parindent=15pt #2\par
	\parindent=15pt #3
	\end{minipage}}\par}}

\renewenvironment{thebibliography}[1]
        {\frenchspacing
	 \ninerm\baselineskip=11pt
         \begin{list}{\arabic{enumi}.}
        {\usecounter{enumi}\setlength{\parsep}{0pt}
	 \setlength{\leftmargin 12.7pt}{\rightmargin 0pt}
         \setlength{\itemsep}{0pt} \settowidth
	{\labelwidth}{#1.}\sloppy}}{\end{list}}

\newcounter{itemlistc}
\newcounter{romanlistc}
\newcounter{alphlistc}
\newcounter{arabiclistc}

\newcommand{\fcaption}[1]{
        \refstepcounter{figure}
        \setbox\@tempboxa = \hbox{\footnotesize Fig.~\thefigure. #1}
        \ifdim \wd\@tempboxa > 5in
           {\begin{center}
        \parbox{5in}{\footnotesize\smalllineskip Fig.~\thefigure. #1}
            \end{center}}
        \else
             {\begin{center}
             {\footnotesize Fig.~\thefigure. #1}
              \end{center}}
        \fi}

\newcommand{\tcaption}[1]{
        \refstepcounter{table}
        \setbox\@tempboxa = \hbox{\footnotesize Table~\thetable. #1}
        \ifdim \wd\@tempboxa > 5in
           {\begin{center}
        \parbox{5in}{\footnotesize\smalllineskip Table~\thetable. #1}
            \end{center}}
        \else
             {\begin{center}
             {\footnotesize Table~\thetable. #1}
              \end{center}}
        \fi}

\def\pmb#1{\setbox0=\hbox{#1}
	\kern-.025em\copy0\kern-\wd0
	\kern.05em\copy0\kern-\wd0
	\kern-.025em\raise.0433em\box0}

\def\fnt#1#2{\footnotetext{\kern-.3em
	{$^{\mbox{\scriptsize #1}}$}{#2}}}

\def\fpage#1{\begingroup
\voffset=.3in
\thispagestyle{empty}\begin{table}[b]\centerline{\footnotesize #1}
	\end{table}\endgroup}

\font\tenit=cmti10
\font\tenbf=cmbx10
\font\bfit=cmbxti10 at 10pt
\font\ninerm=cmr9

\font\eightrm=cmr8

\font\sevenrm=cmr7

\def\FigName{figure}%
\newbox\captionbox
\long\def\@makecaption#1#2{%
  \ifx\FigName\@captype
    \vskip\abovecaptionskip
    \setbox\tempbox\hbox{{\figurecaptionfont #1\hskip1em #2}}
	\ifdim\wd\tempbox< 28pc
	\centerline{\box\tempbox}
	\else
	{\figurecaptionfont #1\hskip1em #2\par}
\fi\else
  	\setbox\tempbox\hbox{{\tablecaptionfont #1\hskip1em #2}}
 	\ifdim\wd\tempbox< 28pc
	\centerline{\box\tempbox}
	\else
	{\tablecaptionfont #1\hskip1em #2\par}%
	\fi
 \vskip\belowcaptionskip
 \fi}
\InputIfFileExists{psfig.sty}
{\typeout{^^Jpsfig.sty inputed...ok}}{\typeout{^^JWarning: psfig.sty could be be found.^^J}}
\InputIfFileExists{epsfsafe.tex}
{\typeout{^^Jepsfsafe.tex inputed...ok}}
			{\typeout{^^JWarning: epsfsafe.tex could not be found.^^J}}
\InputIfFileExists{epsfig.sty}
{\typeout{^^Jepsfig.sty inputed...ok}}{\typeout{^^JWarning: epsfig.sty could not be found.^^J}}
\InputIfFileExists{epsf.sty}
{\typeout{^^Jepsf.sty inputed...ok}}{\typeout{^^JWarning: epsf.sty could not be found.^^J}}%
\def\fps@figure{tbp}
\def\ftype@figure{1}
\def\ext@figure{lof}
\def\fnum@figure{Fig.\ \thefigure}
\usepackage{amsthm}
\usepackage{amsmath}
\usepackage{amssymb}
\usepackage{bbm,amsfonts}
\usepackage{nomencl}
\usepackage{amstext} 

\newcommand{\Eop}{\mathcal{E}}

\newcommand{\Id}{\mathbbm{1}}

\newcommand{\braa}{\langle \psi |}
\newcommand{\kett}{|\psi\rangle}

\newcommand{\Hilb}{\mathcal{H}}

\newcommand{\gatefil}{\mathcal{F}_{\mathcal{E},\mathcal{U}}}

\newcommand{\gatefill}{\mathcal{F}_{\Lambda,\mathcal{I}}}

\newcommand{\gatefilldep}{\mathcal{F}_{\Lambda_{\text{dep}},\mathcal{I}}}
\newcommand{\pure}{\mathbb{CP}^{d-1}}

\newcommand{\Linops}{L\left(\Hilb\right)}

\newtheorem{theorem}{Theorem}

\newtheorem{lemma}{Lemma}
\newtheorem{proposition}{Proposition}
\newtheorem{corollary}{Corollary}

\renewcommand{\thefootnote}{\fnsymbol{footnote}}  

\begin{document}
\setlength{\textheight}{8.0truein}    


\normalsize\textlineskip
\thispagestyle{empty}
\setcounter{page}{1}



\alphfootnote

\fpage{1}

\centerline{\bf
DEPOLARIZING BEHAVIOR OF QUANTUM CHANNELS}
\centerline{\bf IN HIGHER DIMENSIONS}
\vspace*{0.37truein}
\centerline{\footnotesize
EASWAR MAGESAN
}
\vspace*{0.015truein}
\centerline{\footnotesize\it Department of Applied Mathematics and Institute for Quantum Computing, University of Waterloo,}
\centerline{\footnotesize\it 200 University Avenue West, Waterloo, Ontario, Canada N2L 3G1}
\baselineskip=10pt
\vspace*{0.225truein}

\vspace*{0.21truein}

\abstracts{
The paper analyzes the behavior of quantum channels, particularly in large dimensions, by proving various properties of the quantum gate fidelity. Many of these properties are of independent interest in the theory of distance measures on quantum operations. A non-uniqueness result for the gate fidelity is proven, a consequence of which is the existence of non-depolarizing channels that produce a constant gate fidelity on pure states. Asymptotically, the gate fidelity associated with any quantum channel is shown to converge to that of a depolarizing channel. Methods for estimating the minimum of the gate fidelity are also presented.}{}{}

\vspace*{10pt}


\vspace*{1pt}\textlineskip    

\section{Introduction}

Quantum information theory is the study of representing and transforming information using the principles of quantum mechanics. The information is encoded into the set of states for the quantum system and transformed via \emph{quantum operations} (\emph{quantum channels}) which are mathematically represented by completely positive, trace preserving linear maps on the set of states of the system. Thinking about information in this manner has lead to the discovery of quantum algorithms that can solve problems exponentially faster than current classical algorithms~\cite{Sho94,Llo96,HHL}. 
Since quantum operations play a fundamental role in processing and manipulating quantum information, understanding their mathematical properties is of central importance in quantum information theory. One goal of this paper is to understand the behavior of quantum operations in higher dimensions by proving various properties of the \textit{quantum gate fidelity}. In particular, the paper highlights the fact that non-depolarizing quantum channels will exhibit highly depolarizing (isotropic) behavior under certain circumstances.

In many quantum information processing tasks the state of the quantum system is ideally evolved by unitary operations. Experimentally a unitary transformation $\mathcal{U}$ will not be performed perfectly and the actual (implemented) transformation is some general, and likely unknown, quantum operation $\Eop$. A natural question to ask is how distinguishable are $\mathcal{U}$ and $\Eop$ under an appropriate distance measure on quantum channels. The distinguishability of quantum operations has been well-studied in the literature~\cite{Fuc96,Kit97,AKN,GLN}. A comprehensive discussion of various types of distance measures on quantum channels along with an exhaustive set of criteria a useful distance measure should satisfy is given in~\cite{GLN}.

One measure that is particularly useful to use in experimental protocols is the quantum gate fidelity. It can be obtained from the quantum channel fidelity which is a natural extension of the fidelity between quantum states to quantum channels. The channel fidelity between two quantum operations $\Eop_1$ and $\Eop_2$ is the real-valued function on quantum states given by

\begin{equation}
\mathcal{F}_{\Eop_1,\Eop_2}\left(\rho\right)=\left(\text{tr}\sqrt{\sqrt{\mathcal{E}_1(\rho)}\mathcal{E}_2(\rho)\sqrt{\mathcal{E}_1(\rho)}}\right)^2 \nonumber
\end{equation}

\noindent where $\rho$ is an arbitrary mixed quantum state. When the input states are restricted to be pure and the two operations are a unitary $\mathcal{U}$ and a quantum operation $\Eop$, the above function is called the quantum gate fidelity and can be written as,

\begin{equation}
\mathcal{F}_{\Eop,\mathcal{U}}(|\phi\rangle) = \text{tr}\left(\mathcal{U}(|\phi\rangle\langle \phi|)\Eop(|\phi\rangle\langle \phi |)\right) \nonumber
\end{equation}

\noindent for pure state $|\phi\rangle$. The state-dependence of the gate fidelity can be removed by averaging over input states to obtain the \textit{average gate fidelity}, or taking the minimum over all states which gives the \textit{minimum gate fidelity}. These two distance measures satisfy some of the criteria in~\cite{GLN} to be a useful distance measure.


Recently, methods have been given for finding exact expressions of both the average and minimum gate fidelity~\cite{GLN,Nie02,EAZ,PMM,MBKE,KPL} given a description of $\mathcal{U}$ and $\Eop$. An experimental procedure for exactly determining $\Eop$ is given by quantum process tomography~\cite{PCZ,CN97}. Unfortunately since an n quantum bit, or \textit{qubit}, system is represented by a Hilbert space $\Hilb$ of dimension $d=2^n$, process tomography becomes infeasible for even a moderately large number of qubits. As a result, there has recently been interest in providing efficient experimental procedures for characterizing certain features of $\Eop$~\cite{DCEL,ESMR,SMKE,BPP,PMGG}, an example of which is the ability to efficiently estimate the average gate fidelity between $\Eop$ and the identity operation $\mathcal{I}$. Many of the results regarding the quantum gate fidelity in this paper are concerned with its statistical behavior in large dimensions. Another main goal of this paper is to use these results to analyze methods for estimating the average and minimum of the gate fidelity.


\medskip


\noindent The results and structure of the paper are as follows:

\medskip

Section \ref{sec:Background} sets the notation used throughout the paper and presents background material on quantum channels, concentration of measure on the unit sphere, distance measures and the gate fidelity. The main results of the paper are contained in sections \ref{sec:Uniqueness} and \ref{sec:Statistics}.
Section~\ref{sec:Uniqueness} shows that two distinct quantum channels can produce the same gate fidelity function. Specifically if $d \geq 4$ then for any unitary operator $\mathcal{U}$ and full-rank quantum
channel $\Eop_1$ there exists a quantum operation $\Eop_2$ (not equal to either of $\Eop_1$ or $\Eop_1^{\dagger}$) which satisfies 
$\mathcal{F}_{\Eop_1,\mathcal{U}}\left(|\psi\rangle\right) = \mathcal{F}_{\Eop_2,\mathcal{U}}\left(|\psi\rangle\right)$ for every pure state $|\psi\rangle$.
Since depolarizing channels are full-rank, a corollary of this result is that if $d\geq 4$ there exist \textit{non-depolarizing} channels $\Eop$ such that $\mathcal{F}_{\Eop,\mathcal{I}}$ is constant on the set of pure states.



Section \ref{sec:Statistics} analyzes various statistical properties of the gate fidelity, specifically in the large $d$ limit. 
Levy's lemma~\cite{Led01,MS} is used in section \ref{sec:Concentration} to calculate an upper bound on the probability that a randomly chosen state will produce a gate fidelity value
that is far from the average. The measure of the deviating set of states converges to 0 exponentially quickly in the dimension of the quantum system. Section \ref{sec:AveandVariance} uses
these results to obtain upper bounds for the variance of the gate fidelity and
section \ref{sec:Convergence} ties these results together by formalizing the convergence to depolarization of quantum channels. 
Section \ref{sec:Minimum} provides two methods for estimating the minimum gate fidelity using the results of earlier sections. 
The paper concludes in section~\ref{sec:Conclusion} with a discussion of the results and directions for further research.

\section{Background}\label{sec:Background}

This paper will deal only with finite-dimensional quantum systems, therefore quantum systems will be represented by a complex Hilbert space $\Hilb$ of dimension $d < \infty$. The standard isomorphism between $\Hilb$ and $\mathbb{C}^d$ will be assumed without mention throughout the paper. The set of pure states for the system is represented by $\mathbb{C}^d$ modulo phase factors, ie. complex projective space $\pure$. Mixed states for the system are described by the set of positive trace-1 operators in $\Linops$, which will be denoted by $\mathcal{D}(\Hilb)$.

\subsection{Evolution of Quantum Systems and Depolarizing Channels}\label{sec:Evolution}

Let $\Hilb_1$ and $\Hilb_2$ represent finite-dimensional quantum systems of dimensions $d_1$ and $d_2$ respectively. The set of linear superoperators from $L(\Hilb_1)$ to $L(\Hilb_2)$ will be denoted by $\mathcal{T}(\Hilb_1,\Hilb_2)$. A quantum channel, or quantum operation, $\Eop$ is a completely positive, trace-preserving mapping from $L(\Hilb_1)$ into $L(\Hilb_2)$. The set of quantum channels contained in $\mathcal{T}(\Hilb_1,\Hilb_2)$ will be denoted by $\mathcal{S}(\Hilb_1,\Hilb_2)$. Quantum channels describe how an input quantum system is changed under some process or time-evolution. Note that in general the output system of the evolution will be described by a different Hilbert space then the input. In the case that $\Hilb_1=\Hilb_2=\Hilb$, $\mathcal{T}(\Hilb_1,\Hilb_2)$ will be denoted $\mathcal{T}(\Hilb)$ and similarly for $\mathcal{S}(\Hilb_1,\Hilb_2)$.

There are many ways to represent a completely positive, trace-preserving mapping which include the Choi matrix representation~\cite{Choi}, the Kraus representation~\cite{Choi,Kraus} and Stinespring's representation~\cite{Stine}. A good reference for completely positive maps and their representations is given by~\cite{Paul}. We briefly describe the Choi and Kraus representations as they will be used frequently throughout the rest of the presentation.

The Choi matrix for a linear superoperator $\Lambda$ on $L(\Hilb_1)$, denoted $J(\Lambda)$, is the linear operator on $\Hilb_2 \otimes \Hilb_1$ given by,

\begin{equation}
J(\Lambda)= \sum_{(a,b) \in \mathbb{Z}_{d_1}\times \mathbb{Z}_{d_1}}\Lambda(|a\rangle\langle b|)\otimes |a\rangle \langle b|= \left(\Lambda\otimes \mathcal{I}\right)(d_1\sigma)\label{eq:Choimatrixdef}
\end{equation}

\noindent where $\sigma$ is the maximally entangled Bell state state $\left(\frac{1}{\sqrt{D_1}}\sum_{a=1}^{d_1}|a\rangle \otimes |a\rangle \right) \left(\frac{1}{\sqrt{d_1}}\sum_{b=1}^{d_1}\langle b | \otimes \langle b | \right)$. The association $\Lambda \rightarrow J(\Lambda)$ is an isomorphism between $\mathcal{T}(\Hilb_1,\Hilb_2)$ and $L\left(\Hilb_2 \otimes \Hilb_1\right)$. Note also that for any $\Lambda_1$ and $\Lambda_2$, $J(\Lambda_1 \otimes \Lambda_2)=J(\Lambda_1)\otimes J(\Lambda_2)$. From equation (\ref{eq:Choimatrixdef}), $\Lambda$ is completely positive and trace-preserving if and only if $\frac{1}{d_1}J(\Lambda)$ is a quantum state in $L(\Hilb_2 \otimes \Hilb_1)$. Therefore the mapping $\Lambda \rightarrow \frac{1}{d_1}J(\Lambda)$ is a linear isomorphism between quantum operations and quantum states.

A Kraus representation of the linear superoperator $\Lambda$ can be obtained from $J(\Lambda)$. By the singular value decomposition,

\begin{equation}
J(\Lambda)=\sum_{i=1}^k |a_i\rangle\langle b_i| \nonumber
\end{equation}

\noindent where the $|a_i\rangle$ and $|b_i\rangle$ are proportional to the left and right singular vectors of $J(\Lambda)$ respectively, and k is the rank of $J(\Lambda)$. There is an obvious inner-product isomorphism between $L\left(\Hilb_1,\Hilb_2\right)$ with the Hilbert-Schmidt inner product and $\Hilb_2 \otimes \Hilb_1$ with the standard inner product, defined by $|a\rangle\langle b| \rightarrow \text{vec}\left(|a\rangle\langle b|\right) = |a\rangle \otimes |b\rangle$. If $A_i$ and $B_i$ are the unique linear operators in $L\left(\Hilb_1,\Hilb_2\right)$ satisfying $\text{vec}(A_i)=|a_i\rangle$ and $\text{vec}(B_i)=|b_i\rangle$ respectively, then for every $M \in L(\Hilb_1)$,

\begin{equation}
\Lambda(M)=\sum_{i=1}^k A_iM B_i^{\dagger}.
\end{equation}

\noindent The above expression is called a Kraus representation for $\Lambda$ and, unlike the Choi matrix representation, is not unique. If $\Lambda$ is completely positive and trace preserving then $B_i=A_i$ for each i and $\sum_{i=1}^kA_i^{\dagger}A_i = \Id$.




Depolarizing quantum channels on $L\left(\mathbb{C}^d\right)$ are convex combinations of the identity mapping $\mathcal{I}$ and the ``totally depolarizing" mapping $\Omega$ given by

\begin{equation}
\Omega \left(X\right)=\text{tr}\left(X\right)\frac{\Id}{d}.\nonumber
\end{equation}

\noindent Restricting the domain to quantum states implies that a depolarizing channel $\Phi$ has the form,

\begin{equation}
\Phi(\rho)=p\rho + (1-p)\frac{\Id}{d}\nonumber
\end{equation}

\noindent where $p \in [0,1]$ and $\rho$ is an arbitrary quantum state. Clearly $p=1$ corresponds to the identity map $\mathcal{I}$ and $p=0$ corresponds to $\Omega$. 
The set of depolarizing channels in $\mathcal{S}(\Hilb)$ will be denoted by $\mathcal{R}(\Hilb)$.

Sets of Kraus operators for the totally depolarizing channel are given by any unitary 1-design~\cite{DCEL}, examples of which are the generalized Gell-Mann basis~\cite{Geo}, the Heisenberg-Weyl basis and, when $\Hilb=\left({\mathbb{C}^2}\right)^{\otimes n}$, the n-fold tensor product of single qubit Pauli operators. For an excellent discussion of these bases and depolarizing channels see~\cite{Bur09}. Note that all of these bases contain $\Id$ with the remaining operators being traceless. Let $\{\frac{P_i}{d} : i \in \{0,...,d^2-1\}\}$ represent any one of these orthonormal bases with $P_0=\frac{\Id}{d}$. Then,

\begin{equation}
\frac{1}{d^2}\sum_iP_i\rho P_i^{\dagger}=\frac{\Id}{d}\nonumber
\end{equation}

\noindent which gives,

\begin{equation}
\Phi(\rho) = p\rho + \frac{1-p}{d^2}\sum_iP_i\rho P_i^{\dagger} = \left(p+\frac{1-p}{d^2}\right)\rho + \frac{1-p}{d^2}\sum_{i=1}^{d^2}P_i\rho P_i^{\dagger}.\nonumber
\end{equation}

\noindent Therefore the Kraus operators for $\Phi$ are $\sqrt{p+\frac{1-p}{d^2}}\Id$ and $\{\frac{\sqrt{1-p}}{d}P_i:i \in \{1,...,d^2-1\}\}$.

\subsection{Concentration of Measure}\label{sec:BackConcentration}

Concentration of measure, and specifically Levy's lemma, has been utilized in many areas of quantum information to describe the asymptotic behavior of quantum systems in a generic manner. For instance concentration of measure has lead to the proof of the existence of subspaces of bipartite quantum systems consisting entirely of entangled states~\cite{HLW}, explaining thermalization in statistical mechanics~\cite{PSW}, and the construction of counter-examples to the additivity conjecture~\cite{Has09}.

Measure concentration refers to the fact that, on particularly ``concentrated'' metric spaces, slowly varying functions cluster around their median or mean ~\cite{Led01,MS}. The term concentrated is used loosely in the following sense: if one chooses an element A from the Borel algebra of measurable subsets with measure $\geq \frac{1}{2}$ then, for any $\epsilon > 0$, the $\epsilon$-neighbourhood of A has measure close to 1. One well known example of a space that exhibits this property is the unit sphere $\mathbb{S}^n \subset \mathbb{R}^{n+1}$.

Suppose $\mathbb{S}^n$ is endowed with the Euclidean metric $\|\: \|_2$. A function $f : \mathbb{S}^n \rightarrow \mathbb{R}$ is called $K$-Lipschitz if $\forall \: x,y \in \mathbb{S}^n$,

\begin{equation}
|f(x)-f(y)| \leq K\|x-y\|_2.\nonumber
\end{equation}

\noindent  Let $``f \in (a,b)"$ be notation for the set of all points in $\mathbb{S}^n$ whose image under f lies in $(a,b)$ and let $\mu$ be the rotationally invariant Haar probability measure on $\mathbb{S}^n$. From \cite{MS,HLW} Levy's lemma states that if $f : \mathbb{S}^n \rightarrow \mathbb{R}$ is $K$-Lipschitz then,

\begin{equation}
\mathbb{P}_{\mu} \left[f \in \left(-\infty, \mathbb{E}_{\mu}\left[f\right]-\epsilon\right)\right] \leq 2e^{\frac{-C_1 \epsilon ^2 (n+1)}{K ^2}}.\nonumber
\end{equation}

\noindent where $\mathbb{E}_{\mu}\left[f\right] = \int f d\mu$ and the constant $C_1$ can be taken to be $\frac{1}{9\pi ^3 ln 2}$. An analogous inequality holds for the interval $\left(\mathbb{E}_{\mu}\left[f\right] + \epsilon, \infty \right)$ which implies,

\begin{equation}
\mathbb{P}_{\mu} \left[f \in \left(\mathbb{E}_{\mu}\left[f\right]-\epsilon, \mathbb{E}_{\mu}\left[f\right] + \epsilon \right)\right] \geq 1-4e^{\frac{-C_1 \epsilon ^2 (n+1)}{K ^2}}.\nonumber
\end{equation}

\noindent The above statement reads that if x is chosen uniformly at random according to $\mu$ then the probability f(x) lies in the interval $\left(\mathbb{E}_{\mu}\left[f\right]-\epsilon, \mathbb{E}_{\mu}\left[f\right] + \epsilon \right)$ is greater than or equal to $1-4e^{\frac{-C_1 \epsilon ^2 (n+1)}{K ^2}}$. From~\cite{MS} equivalent inequalities hold for the median of f.

Levy's lemma for the real unit sphere can be translated into results relevant to quantum theory. Analogous to the Haar measure on $\mathbb{S}^{2d-1}$, the Borel measure induced by the Fubini-Study metric~\cite{BZ} on $\mathbb{CP}^{d-1}$ is the unique unitarily invariant probability measure on $\mathbb{CP}^{d-1}$. This measure is called the Fubini-Study measure and will be denoted $\mu_{F}$.

Any function g from $\pure$ into $\mathbb{R}$ can be thought of as a function from the set of unit vectors in $\mathbb{C}^d$, denoted $\mathbb{S}^{\mathbb{C}^d}$, into $\mathbb{R}$ that is independent of the relative phase between vectors. By the obvious isomorphism between $\mathbb{S}^{\mathbb{C}^d}$ and $\mathbb{S}^{2d-1}$, g can equivalently be thought of as a function h from $\mathbb{S}^{2d-1}$ into $\mathbb{R}$. Moreover, if g is integrable with respect to $\mu_{F}$ on $\pure$,

\begin{equation}
\mathbb{E}_{\mu_{F}}\left[g\right]=\mathbb{E}_{\mu}\left[h\right].\nonumber
\end{equation}

\noindent If $\alpha=\mathbb{E}_{\mu_{F}}\left[g\right]$,

\begin{equation}
\mathbb{P}_{\mu_{F}} \left[g \in \left(\alpha-\epsilon, \alpha +\epsilon\right)\right] = \mathbb{P}_{\mu} \left[h \in \left(\alpha-\epsilon, \alpha+\epsilon\right)\right]\nonumber,
\end{equation}

\noindent and so the concentration inequalities given above for the real unit sphere can be translated to $\pure$ equipped with the Fubini-Study measure.

\subsection{The Quantum Gate Fidelity and Distance Measures}\label{sec:gatefilanddist}

The fidelity F between $\rho$ and $\sigma$ in $\mathcal{D}(\Hilb)$ is defined by,

\begin{equation}
F\left(\rho,\sigma\right) = \left(\text{tr}\sqrt{\sqrt{\rho}\sigma\sqrt{\rho}}\right)^2.\nonumber
\end{equation}

\noindent F is a useful measure of how far apart two states are in terms of deviation of measurement statistics~\cite{NC}. The gate-fidelity $\mathcal{F}_{\Eop,\mathcal{U}}$ is a state-dependent description of the distance between the unitary $\mathcal{U}$ and $\Eop \in \mathcal{S}(\Hilb)$. If $\rho \in \mathcal{D}(\mathcal{H})$ then $\mathcal{F}_{\Eop,\mathcal{U}}$ is defined as,

\begin{equation}
\mathcal{F}_{\Eop,\mathcal{U}}(\rho) := F\left(\mathcal{E}(\rho),\mathcal{U}(\rho)\right) = \left(\text{tr}\sqrt{\sqrt{\mathcal{E}(\rho)}\mathcal{U}(\rho)\sqrt{\mathcal{E}(\rho)}}\right)^2.\nonumber
\end{equation}

For this paper the case of interest is when the input state is pure. If $|\phi\rangle \in \mathbb{CP}^{d-1}$,

\begin{equation}
\mathcal{F}_{\Eop,\mathcal{U}}(|\phi\rangle) = \text{tr}\left(\mathcal{U}(|\phi\rangle\langle \phi|)\Eop(|\phi\rangle\langle \phi |)\right)
\end{equation}

\noindent and if $\{M_k\}$ and $U$ are Kraus operators for $\Eop$ and $\mathcal{U}$ respectively,

\begin{equation}
\mathcal{F}_{\Eop,\mathcal{U}}(|\phi\rangle) = \text{tr}\left(U|\phi\rangle\langle \phi | U^{\dagger} \sum_k M_k |\phi\rangle\langle \phi | M_k^{\dagger}\right) = \text{tr}\left(|\phi\rangle\langle \phi | \: \mathcal{U}^{\dagger} \circ \Eop (|\phi\rangle\langle \phi |)\right).\nonumber
\end{equation}

\noindent Defining $\Lambda = \mathcal{U}^{\dagger} \circ \Eop$,

\begin{equation}
\mathcal{F}_{\Eop,\mathcal{U}}(|\phi\rangle) = \text{tr}\left(|\phi\rangle\langle \phi | \Lambda (|\phi\rangle\langle \phi |)\right) = \gatefill(|\phi\rangle).\label{eq:EoptoLambda}
\end{equation}

\noindent $\Lambda$ is a quantum operation that, loosely speaking, is a measure of how much $\Eop$ deviates from $\mathcal{U}$. From (\ref{eq:EoptoLambda}), many of the results that will be proved for $\gatefil$ will without loss of generality be proven for $\gatefill$.

The following result can be easily proven and will be used later,

\bigskip

\begin{proposition}\label{thm:Depconst}

If $\Eop$ is depolarizing with $\Eop (\rho) = p\rho + (1-p)\frac{\Id}{d}$ and $\mathcal{U}=\mathcal{I}$ then for every pure state $|\phi\rangle$,

\begin{equation}
\gatefil (|\phi\rangle) = p+\frac{1-p}{d}.\nonumber
\end{equation}

\end{proposition}

\bigskip

\noindent Hence the gate fidelity between a depolarizing channel and the identity operation is constant on $\pure$.

Two important measures of distance between $\Eop$ and $\mathcal{U}$ derived from $\gatefil$ are the average of $\gatefil$ and the minimum of $\gatefil$. The average, $\mathbb{E}_{\mu_F}\left[\mathcal{F}_{\Eop,\mathcal{U}}\right]$, is given by,

\begin{align}
\mathbb{E}_{\mu_F}\left[\mathcal{F}_{\Eop,\mathcal{U}}\right] &= \int_{\pure} \text{tr}\left(\kett \braa \Lambda (\kett \braa)\right) d\mu_F\left(\psi\right) \nonumber\\
&= \frac{\sum _i \left( \text{tr}(K_i)\text{tr}(K_i^{\dagger})\right) + 1}{d^2 + d}\nonumber
\end{align}

\noindent where the $\{K_i\}$ are a set of Kraus operators for $\Lambda$~\cite{Nie02,EAZ}. For the rest of the paper, the more common notation of $\overline{\gatefil}$ will be used instead of $\mathbb{E}_{\mu_F}\left[\mathcal{F}_{\Eop,\mathcal{U}}\right]$. $\overline{\gatefil}$ is useful because it is a single parameter describing the distance between $\mathcal{U}$ and $\Eop$. The minimum of the gate fidelity over $\pure$, $\mathcal{F}_{\Eop,\mathcal{U}}^{\text{min}}$, is also of interest because it characterizes the worst case fidelity between the outputs of $\mathcal{U}$ and $\Eop$. By concavity of the fidelity~\cite{NC}, the minimum over pure states is equal to the minimum over all mixed quantum states.

Six properties that a useful measure of distance, $\Delta$, should satisfy are discussed in~\cite{GLN} and listed here for reference,

\bigskip

\noindent 1. \textit{Metric}: $\Delta$ should be a metric.

\smallskip

\noindent 2. \textit{Easy to calculate}: There should be a straightforward method for evaluating $\Delta$.

\smallskip

\noindent 3. \textit{Easy to measure}: There should be a clear and achievable experimental protocol for determining $\Delta$.

\smallskip

\noindent 4. \textit{Physical interpretation}: $\Delta$ should have a well-motivated physical interpretation

\smallskip

\noindent 5. \textit{Stability}: $\Delta$ should be stable under tensoring with the identity operation, ie. if $\mathcal{Q}$ and $\mathcal{R}$ are quantum operations, $\Delta\left(\mathcal{Q}\otimes \mathcal{I}, \mathcal{R}\otimes \mathcal{I}\right) = \Delta\left(\mathcal{Q},\mathcal{R}\right)$.

\smallskip

\noindent 6. \textit{Chaining}: For a process composed of many smaller steps, the total error will be less than the sum of the errors in the individual steps, ie. for channels $\mathcal{Q}_1$, $\mathcal{Q}_2$, $\mathcal{R}_1$ and $\mathcal{R}_2$, $\Delta(\mathcal{Q}_2\circ \mathcal{Q}_1, \mathcal{R}_2 \circ \mathcal{R}_1) \leq \Delta(\mathcal{Q}_2,\mathcal{R}_2) + \Delta(\mathcal{Q}_1,\mathcal{R}_1)$.

\bigskip

$\overline {\mathcal{F}_{\Eop,\mathcal{U}}}$ and$\mathcal{F}_{\Eop,\mathcal{U}}^{\text{min}}$ are both candidates to be a good measure of distance. $\overline {\mathcal{F}_{\Eop,\mathcal{U}}}$ is shown in~\cite{GLN} to satisfy properties 2, 3 and 4 but fails to satisfy the rest. $\mathcal{F}_{\Eop,\mathcal{U}}^{\text{min}}$ on the other hand satisfies all of the properties except for 2 and 3. It should be noted that if process tomography can be performed then $\mathcal{F}_{\Eop,\mathcal{U}}^{\text{min}}$ can be calculated numerically using convex optimization techniques.

%





\section{Non-Uniqueness of the Gate Fidelity}\label{sec:Uniqueness}

As mentioned in the introduction, the gate fidelity is particularly important in experimental quantum computation because the ideal transformation is a unitary superoperator, while the implemented (real) transformation is some general quantum operation. A question that arises is, if the intended unitary operation is $\mathcal{U}$, then does the gate fidelity on $\pure$ uniquely characterize the implemented quantum operation? Equivalently, if the unitary operator $\mathcal{U}$ is fixed then can there exist two distinct quantum channels $\mathcal{Q}$ and $\mathcal{R}$ satisfying $\mathcal{F}_{\mathcal{Q},\mathcal{U}} = \mathcal{F}_{\mathcal{R},\mathcal{U}}$? From (\ref{eq:EoptoLambda}) this question is equivalent to the problem of determining whether there exist two distinct quantum channels $\mathcal{Q}$ and $\mathcal{R}$ such that $\mathcal{F}_{\mathcal{Q},\mathcal{I}} = \mathcal{F}_{\mathcal{R},\mathcal{I}}$.

It is clear that the gate fidelity is not unique in general by noting that if $\Eop$ is a channel such that $\Eop \neq \Eop^{\dagger}$ then

\begin{equation}
\text{tr}\left(\Eop(\kett\braa)\kett\braa\right)=\text{tr}\left(\kett\braa \Eop^{\dagger}(\kett\braa)\right). \nonumber
\end{equation}




\noindent The main theorem of this section shows that if $d\geq 4$ and $\mathcal{Q}$ is a full-rank quantum operation then there exists a quantum channel $\mathcal{R} \neq \mathcal{Q}^{\dagger}$ which produces the \textit{same} gate fidelity function. In this context, full-rank means that the minimum number of Kraus operators required for $\mathcal{Q}$ is $d^2$. From section \ref{sec:Evolution} this requirement is equivalent to the Choi matrix of $\mathcal{Q}$ being positive definite.

\bigskip

\begin{theorem}\label{thm:nonuniqueness}

Suppose that dim($\Hilb$)$= d \geq 4$ and $\mathcal{Q}$ is a quantum operation on $\Linops$ with a positive-definite Choi matrix. Then there exists a quantum channel $\mathcal{R} \neq \mathcal{Q}^{\dagger}$ (and $\mathcal{R} \neq \mathcal{Q}$) such that

\begin{equation}
\mathcal{F}_{\mathcal{Q},\mathcal{I}} = \mathcal{F}_{\mathcal{R},\mathcal{I}}.\nonumber
\end{equation}

\end{theorem}


\bigskip

In order to prove theorem \ref{thm:nonuniqueness} we will need the following lemma:

\bigskip

\begin{lemma}\label{lemma:sufficient}

A linear superoperator $\Lambda$ acting on $\Linops$ can be written as the difference between two quantum operations $\Lambda_1$ and $\Lambda_2$ satisfying $\mathcal{F}_{\Lambda_1,\mathcal{I}} = \mathcal{F}_{\Lambda_2,\mathcal{I}}$ if the following conditions are satisfied,

\bigskip

1. $J(\Lambda)$ is the difference between two positive semi-definite operators A and B such that $\text{tr}_{\Hilb_1}A= \text{tr}_{\Hilb_1}B=\Id$,

\bigskip

2. If $\mathcal{I} \otimes T$ represents the partial transpose operation on L($\Hilb_1 \otimes \Hilb_2$) then $\left(\mathcal{I} \otimes T\right)\left(J(\Lambda)\right)$ has support on the anti-symmetric subspace of $\Hilb_1 \otimes \Hilb_2$.

\bigskip

\begin{proof}(Lemma)

First, suppose that $J(\Lambda)$ is equal to $A-B$ where $A$ and $B$ are positive semi-definite operators and $\text{tr}_{\Hilb_1}A= \text{tr}_{\Hilb_1}B=\Id$. From section \ref{sec:Evolution} these assumptions on A and B are equivalent to $A=J(\Lambda_1)$ and $B=J(\Lambda_2)$ for quantum operations $\Lambda_1$ and $\Lambda_2$. Thus by linearity, condition 1 is equivalent to $\Lambda=\Lambda_1-\Lambda_2$ where $\Lambda_1$ and $\Lambda_2$ are quantum operations. Hence it remains to show that the second condition implies $\mathcal{F}_{\Lambda_1,\mathcal{I}} = \mathcal{F}_{\Lambda_2,\mathcal{I}}$.

Since the vec correspondence between $\Linops$ with the Hilbert-Schmidt inner product and $\Hilb_1 \otimes \Hilb_2$ with the standard inner product is an inner-product isomorphism (see section \ref{sec:Evolution}), for any A,B in $\Linops$,

\begin{equation}
\langle A,B\rangle = \text{tr}\left(A^{\dagger}B\right) = \text{vec}(A)^{\dagger}\text{vec}(B)=\langle \text{vec}(A),\text{vec}(B)\rangle.\nonumber
\end{equation}

\noindent If $J(\Lambda)$ has spectral decomposition,

\begin{equation}
J(\Lambda)= \sum_i\lambda_i \text{vec}(A_i)\text{vec}(A_i)^{\dagger},\nonumber
\end{equation}

\noindent then,

\begin{equation}
\langle J(\Lambda), |m\rangle \otimes |n\rangle \langle k |\otimes \langle l|\rangle = \sum_i \lambda_i\langle |k\rangle \otimes |l\rangle , \text{vec}(A_i)\rangle \langle \text{vec}(A_i), |m\rangle \otimes |n\rangle \rangle. \nonumber
\end{equation}

\noindent The vec correspondence again gives,

\begin{eqnarray*}
\sum_i \lambda_i\langle |k\rangle \otimes |l\rangle , \text{vec}(A_i)\rangle \langle \text{vec}(A_i), |m\rangle \otimes |n\rangle \rangle &=& \sum_i \lambda_i \text{tr}\left( \left(|k\rangle \langle l |\right)^{\dagger} A_i\right) \text{tr}\left( A_i^{\dagger} |m\rangle \langle n|\right) \\ \nonumber
&=& \text{tr}\left( \Lambda\left(|l\rangle \langle n|\right)|m\rangle \langle k|\right)\nonumber
\end{eqnarray*}

\noindent and so,

\begin{equation}
\langle J(\Lambda), |m\rangle \otimes |n\rangle \langle k |\otimes \langle l|\rangle=\text{tr}\left( \Lambda\left(|l\rangle \langle n|\right)|m\rangle \langle k|\right).\label{eq:innerpeq}
\end{equation}




\noindent Noting that,

\begin{equation}
\text{tr}\left( \Lambda\left(|l\rangle \langle n|\right)|m\rangle \langle k|\right) = \text{tr}\left(J(\Lambda) \left[|m\rangle \langle k | \otimes \left(|l\rangle \langle n |\right)^T\right] \right) = \text{tr}\left(J(\Lambda) \left[\mathcal{I} \otimes T \left(|m\rangle \langle k | \otimes |l\rangle \langle n | \right)\right]\right) \nonumber
\end{equation}


\noindent and,

\begin{equation}
\text{tr}\left(J(\Lambda) \left[\mathcal{I} \otimes T \left(|m\rangle \langle k | \otimes |l\rangle \langle n | \right)\right]\right) = \text{tr} \left(\left[\mathcal{I} \otimes T\left(J(\Lambda)\right)\right] |m\rangle \langle k | \otimes |l\rangle \langle n | \right), \nonumber
\end{equation}

\noindent for any $\kett \in \pure$,

\begin{equation}
\text{tr}\left(\Lambda(\kett\braa)\kett\braa\right)=\text{tr}\left(\left[\mathcal{I} \otimes T\left(J(\Lambda)\right)\right]\kett\braa \otimes \kett\braa\right).\nonumber
\end{equation}

\noindent Hence $\text{tr}\left(\Lambda(\kett\braa)\kett\braa\right)=0$ if and only if $\text{tr}\left(\left[\mathcal{I} \otimes T\left(J(\Lambda)\right)\right]\kett\braa \otimes \kett\braa\right)=0$.

\bigskip

In total, the above discussion shows that the conditions:

\bigskip

1. $J(\Lambda)$ is the difference between two positive semi-definite operators A and B such that $\text{tr}_{\Hilb_1}A= \text{tr}_{\Hilb_1}B=\Id$,

\bigskip

2. For every $\kett\braa$, $\text{tr}\left(\left[\mathcal{I} \otimes T\left(J(\Lambda)\right)\right]\kett\braa \otimes \kett\braa\right)=0$,

\bigskip

\noindent are satisfied if and only if $\Lambda$ can be written as the difference between two quantum operations $\Lambda_1$ and $\Lambda_2$ satisfying $\mathcal{F}_{\Lambda_1,\mathcal{I}} = \mathcal{F}_{\Lambda_2,\mathcal{I}}$.

Let the symmetric and anti-symmetric subspace in $\Hilb_1\otimes \Hilb_2$ be denoted sym(2,$d$) and \text{a-sym}(2,$d$) respectively so that $\Hilb_1 \otimes \Hilb_2 = \text{sym}(2,d) \oplus \text{a-sym}(2,d)$. Since every state $\kett$ satisfies $\kett \otimes \kett \in \text{sym}(2,d)$, if $\left(\mathcal{I} \otimes T\right)\left(J(\Lambda)\right)$ has support on $\text{a-sym}(2,d)$ then

\begin{equation}
\text{tr}\left(\Lambda(\kett\braa)\kett\braa\right)=\text{tr}\left(\left[\left(\mathcal{I} \otimes T\right)\left(J(\Lambda)\right)\right]\kett\braa \otimes \kett\braa\right)=0 \nonumber
\end{equation}

\noindent for every $\kett$. Thus the conditions:

\bigskip

1. $J(\Lambda)$ is the difference between two positive semi-definite operators A and B such that $\text{tr}_{\Hilb_1}A= \text{tr}_{\Hilb_1}B=\Id$,

\bigskip

2. $\left(\mathcal{I} \otimes T\right)\left(J(\Lambda)\right)$ has support on a-sym(2,d),

\bigskip

\noindent are sufficient for $\Lambda$ to be the difference between two quantum operations which produce the same gate-fidelity.

\end{proof}

\end{lemma}

\bigskip

Theorem \ref{thm:nonuniqueness} can now be proven using lemma \ref{lemma:sufficient}.

\begin{proof} (Theorem)

First, let $d=4$ so that $\Hilb_1$ and $\Hilb_2$ are both identified with $\mathbb{C}^4$ and suppose $\mathcal{Q}$ is such that $J(\mathcal{Q}) > 0$. $\mathcal{R}$ is explicitly constructed by first showing that there is an element of $L(\Hilb_1\otimes \Hilb_2)$ satisfying the two conditions from lemma \ref{lemma:sufficient}. Define,

\begin{gather}
|\alpha_1\rangle = \frac{1}{\sqrt{2}}(|01\rangle - |10\rangle), \: \: |\beta_1\rangle = \frac{1}{\sqrt{2}}(|23\rangle - |32\rangle),\nonumber \\
|\alpha_2\rangle = \frac{1}{\sqrt{2}}(|02\rangle - |20\rangle), \: \: |\beta_2\rangle = \frac{1}{\sqrt{2}}(|13\rangle - |31\rangle), \nonumber \\
|\alpha_3\rangle = \frac{1}{\sqrt{2}}(|03\rangle - |30\rangle), \: \: |\beta_3\rangle = \frac{1}{\sqrt{2}}(|12\rangle - |21\rangle). \nonumber
\end{gather}

\noindent These six vectors form an orthonormal basis for a-sym(2,4). Define G $\in L(\Hilb_1\otimes \Hilb_2)$ via the equation,

\begin{equation}
\left(\mathcal{I}\otimes T\right)(G)=|\alpha_1\rangle\langle\beta_1| + |\alpha_2\rangle\langle\beta_2| + |\alpha_3\rangle\langle\beta_3|+ |\beta_1\rangle\langle\alpha_1| + |\beta_2\rangle\langle\alpha_2| + |\beta_3\rangle\langle\alpha_3|.\nonumber
\end{equation}

\noindent It is straightforward to verify that G is Hermitian, $\text{tr}_{\Hilb_1}(G)=\text{tr}_{\Hilb_1}(\left(\mathcal{I}\otimes T\right)(G))=0$ and $\left(\mathcal{I}\otimes T\right)(G)$ has support on $\text{a-sym}(2,4)$.

Let $\mathcal{G}$ be the unique linear superoperator such that $J(\mathcal{G})=G$. Since $J(\mathcal{Q})>0$ there exists $\epsilon > 0$ depending on both $\mathcal{Q}$ and $\mathcal{G}$ such that

\begin{equation}
J(\mathcal{Q}) + \epsilon J(\mathcal{G}) \geq 0. \nonumber
\end{equation}

\noindent Thus $\epsilon \mathcal{G}$ is such that,

\bigskip

1. $J(\epsilon\mathcal{G})=J(\mathcal{Q}+\epsilon\mathcal{G})-J(\mathcal{Q})$ with $J(\mathcal{Q})$, $J(\mathcal{Q}+\epsilon\mathcal{G}) \geq 0$ and $\text{tr}_{\Hilb_1}J(\mathcal{Q}+\epsilon\mathcal{G})=\text{tr}_{\Hilb_1}J(\mathcal{Q})=\Id$,

\bigskip

2. $\left(\mathcal{I}\otimes T\right) \left(J(\epsilon\mathcal{G})\right) = \epsilon \left(\mathcal{I}\otimes T\right)(G)$ has support on a-sym(2,d).

\bigskip

\noindent Hence from lemma \ref{lemma:sufficient}, $\mathcal{Q}$ and $\mathcal{R}:=\mathcal{Q}+\epsilon\mathcal{G}$ are two quantum operations that produce the same gate fidelity. Up to finding an explicit value for $\epsilon$ this proves the theorem for $d=4$.

To find a value for $\epsilon$ note that since $J(\mathcal{Q}) > 0$, the smallest eigenvalue of $J(\mathcal{Q})$, denoted $\lambda_{\text{min}}^{\mathcal{Q}}$, is strictly greater than 0. Therefore for every vector $|\phi\rangle \in \mathbbm{C}^4 \otimes \mathbbm{C}^4$,

\begin{equation}
\langle \phi |J(\mathcal{Q})|\phi\rangle \in \left[\lambda_{\text{min}}^{\mathcal{Q}}, \|J(\mathcal{Q})\|_{\infty}\right] \nonumber
\end{equation}

\noindent Moreover, since $\langle \phi |J(\mathcal{G})|\phi\rangle \in \left[-\epsilon\|J(\mathcal{G})\|_{\infty}, \epsilon\|J(\mathcal{G})\|_{\infty}\right]$,

\begin{equation}
\langle \phi |J(\mathcal{Q}+\epsilon\mathcal{G})|\phi\rangle \in \left[\lambda_{\text{min}}^{\mathcal{Q}}-\epsilon\|J(\mathcal{G})\|_{\infty}, \|J(\mathcal{Q})\|_{\infty} + \epsilon \|J(\mathcal{G})\|_{\infty}\right].\nonumber
\end{equation}

\noindent Therefore in order for $J(\mathcal{Q}+\epsilon\mathcal{G}) \geq 0$ to be satisfied it must be that

\begin{equation}
0<\epsilon \leq \frac{\lambda_{\text{min}}^{\mathcal{Q}}}{\|J(\mathcal{G})\|_{\infty}}.\nonumber
\end{equation}

Lastly, suppose $d > 4$. Since the vector space spanned by $\{|\alpha_1\rangle,|\alpha_2\rangle,|\alpha_3\rangle,|\beta_1\rangle,|\beta_2\rangle,|\beta_3\rangle\}$ is a subspace of a-sym(2,d), $\mathcal{Q}+\epsilon \mathcal{G}$ can be defined in the same manner as above which proves the theorem.

\end{proof}

\bigskip

The following corollary follows immediately from theorem \ref{thm:nonuniqueness}.

\bigskip

\begin{corollary} \label{cor:nondepconstant}

Let $\text{dim}\left(\Hilb\right) = d\geq 4$. Suppose $\mathcal{Q}$ is a depolarizing channel on $L(\Hilb)$ of the form

\begin{gather}
\mathcal{Q} (A)=pA + (1-p)\text{tr}(A)\frac{\Id}{d} \nonumber
\end{gather}

\noindent where $p \in [0,1)$ and let $\mathcal{G}$ be the linear superoperator from theorem \ref{thm:nonuniqueness}. Then for any $\epsilon \in \left(0,\frac{1-p}{d\|J(\mathcal{G})\|_{\infty}}\right]$, $\mathcal{R} = \mathcal{Q} + \epsilon \mathcal{G}$ is a non-depolarizing quantum operation with $\mathcal{F}_{\mathcal{Q},\mathcal{I}}=\mathcal{F}_{\mathcal{R},\mathcal{I}}$.

\bigskip

\begin{proof}

Since $\mathcal{Q}$ is depolarizing with $p \in [0,1)$, $J\left(\mathcal{Q}\right)$ is a positive matrix. Thus \ref{thm:nonuniqueness} gives both the existence and construction of $\mathcal{R}$ in terms of $\mathcal{G}$. The fact that $\epsilon$ lies in $\left(0,\frac{1-p}{d\|J(\mathcal{G})\|_{\infty}}\right]$ follows from the fact that $\lambda_{\text{min}}^{\mathcal{Q}} = \frac{1-p}{d}$.

\end{proof}

\end{corollary}

\bigskip

Corollary \ref{cor:nondepconstant} shows that the gate fidelity cannot always distinguish between depolarizing and non-depolarizing quantum channels. The following is a straightforward result of proposition \ref{thm:Depconst} and corollary \ref{cor:nondepconstant},

\bigskip

\begin{corollary} \label{cor:nondepconstant1}

There exist non-depolarizing quantum channels $\Eop$ such that $\mathcal{F}_{\Eop,\mathcal{I}}$ is constant on $\pure$.

\end{corollary}

\bigskip

In terms of the Bloch representation of quantum states~\cite{Blo46,RSW}, the action of a depolarizing channel is to isotropically shrink the Bloch object. Corollary \ref{cor:nondepconstant1} shows that even if the gate fidelity between $\Eop$ and $\mathcal{I}$ is a constant function, one is unable to deduce whether $\Eop$ isotropically shrinks the Bloch object.

%






\section{Statistical Properties and Asymptotic Behavior of the Gate Fidelity}\label{sec:Statistics}

The aim of this section is to deduce various statistical properties of the gate fidelity, many of which are asymptotic. This is done by viewing the gate fidelity as a random variable on $\pure$, where we assume $\pure$ is equipped with the Fubini-Study measure $\mu_F$. By equation (\ref{eq:EoptoLambda}) there is no loss in generality in restricting attention to gate fidelities of the form $\gatefill$ where $\Lambda$ is some quantum operation.

The variance of $\gatefill$, denoted $\sigma^2(\Lambda)$, is given by

\begin{equation}
\sigma^2(\Lambda)=\mathbb{E}_{\mu_F}\left[\left(\gatefill - \overline{\gatefill}\right)^2\right]= \overline{\mathcal{F}_{\Lambda,\mathcal{I}}^2}-\overline{\mathcal{F}_{\Lambda,\mathcal{I}}}^2. \nonumber 
\end{equation}

\noindent If $\Lambda$ is depolarizing then $\sigma^2(\Lambda)=0$ and from the previous section, for d $\geq 4$, a non-depolarizing quantum channel $\mathcal{R}$ was constructed which satisfies $\mathcal{F}_{\mathcal{R},\mathcal{I}}=\gatefill$. Hence there exists non-depolarizing quantum channels $\mathcal{R}$ with $\sigma^2(\mathcal{R})=0$. Therefore it is \textit{not} true that $\Lambda$ is a depolarizing channel if and only if the variance of $\gatefill$ is 0.

From~\cite{MBKE},

\begin{equation}
\sigma^2(\Lambda) = \frac{\overline{\mathcal{F}_{\Lambda,\mathcal{I}}}^2 d^4 + O(d^3)}{d^4 + 6d^3 + 11d^2 + 6d} - \overline{\mathcal{F}_{\Lambda,\mathcal{I}}}^2 \sim O\left(\frac{1}{d}\right)\nonumber
\end{equation}

\noindent and so $\sigma^2(\Lambda) \rightarrow 0$ as $\frac{1}{d}$ when $d \rightarrow \infty$. In fact, an explicit upper bound for $\sigma^2(\Lambda)$ is given by,

\begin{equation}
\sigma^2(\Lambda) \leq \frac{8d^3 + 16d^2 + 4d}{\left(d^2+2d+1\right)\left(d^2+5d+1\right)}. \label{eq:varupperbound}
\end{equation}

\noindent Equation \ref{eq:varupperbound} holds for \textit{any} quantum channel. Therefore for large d and any channel $\Lambda$, the second central moment of $\gatefill$ is very small. This implies that $\gatefill$ must be ``close" to $\gatefilldep$ as random variables, which will be made precise in section \ref{sec:Convergence} using both a natural metric on $\xi$ and bounds obtained in section \ref{sec:Concentration}.

\subsection{Concentration of Measure for the Gate Fidelity}\label{sec:Concentration}

In this section, Levy's lemma (discussed in section \ref{sec:BackConcentration}) is used to make precise the idea that $\gatefill(|\phi\rangle)$ is close to $\overline {\gatefill}$ when $|\phi\rangle$ is chosen uniformly at random according to the Fubini-Study measure. The key is to show that $\gatefill$ satisfies a Lipschitz condition which is \textit{independent} of the dimension d
of the system.

\bigskip

\begin{theorem}

The function $\gatefill : (\mathbb{CP}^{d-1}, \|\:\|_2) \rightarrow [0,1]$ satisfies a K-Lipschitz condition for some $K \geq 0$ independent of d.

\bigskip

\begin{proof}

\bigskip

The goal is to show that $\forall |\phi_1\rangle, |\phi_2\rangle \in \mathbb{CP}^{d-1}$,

\begin{equation}
|\gatefill(|\phi_1\rangle) - \gatefill(|\phi_2\rangle)| \leq K\||\phi_1\rangle-|\phi_2\rangle\|_2, \nonumber
\end{equation}

\noindent where K is independent of d. 
By the triangle inequality,

\begin{eqnarray}
|\gatefill(|\phi_1\rangle) - \gatefill(|\phi_2\rangle)| &\leq&|\text{tr}\left(|\phi _1\rangle \langle \phi _1|\left(\Lambda \left(|\phi _1\rangle \langle \phi _1|\right)-\Lambda \left(|\phi _2\rangle \langle \phi _2|\right)\right)\right)| \nonumber \\
& \: &  + |\text{tr}\left(\Lambda (|\phi _2\rangle \langle \phi _2|)\left(|\phi _1\rangle \langle \phi _1|-|\phi _2\rangle \langle \phi _2|\right)\right)|.\nonumber
\end{eqnarray}

\noindent Let $\| \: \|_1$ and $\| \: \|_2$ be the Schatten 1 and 2-norms (ie. trace and Frobenius norms) on $L\left(\Hilb\right)$ respectively~\cite{Wat05}. By the Cauchy-Schwarz inequality,

\begin{eqnarray}
|\gatefill(|\phi_1\rangle) - \gatefill(|\phi_2\rangle)| &\leq& \| |\phi _1\rangle \langle \phi _1|\|_2 \| \Lambda (|\phi _1\rangle \langle \phi _1| - |\phi _2\rangle \langle \phi _2|)\|_2 \nonumber \\
&\:& + \| \Lambda (|\phi _2\rangle \langle \phi _2|)\|_2 \| |\phi _1\rangle \langle \phi _1| - |\phi _2\rangle \langle \phi _2|\|_2.\nonumber
\end{eqnarray}

For any linear operator $A \in L\left(\Hilb\right)$ \cite{HJ},

\begin{equation}
\|A\|_2 \leq \|A\|_1 \leq \text{rank}(A)\|A\|_2\nonumber
\end{equation}

\noindent which gives $\| \Lambda (|\phi _2\rangle \langle \phi _2|)\|_2 \leq \| \Lambda (|\phi _2\rangle \langle \phi _2|)\|_1 = 1$. As well for any pure state $|\psi\rangle$, $\| |\psi \rangle \langle \psi| \|_2 = 1$. Therefore,

\begin{equation}
|\gatefill(|\phi_1\rangle) - \gatefill(|\phi_2\rangle)| \leq \| \Lambda (|\phi _1\rangle \langle \phi _1| - |\phi _2\rangle \langle \phi _2|)\|_1 + \| |\phi _1\rangle \langle \phi _1| - |\phi _2\rangle \langle \phi _2|\|_2.\nonumber
\end{equation}

\noindent  Using the fact that quantum operations can only decrease the $\| \: \|_1$ distance between quantum states~\cite{NC} and also that the difference of two rank 1 projectors has rank at most 2,

\begin{equation}
|\gatefill(|\phi_1\rangle) - \gatefill(|\phi_2\rangle)| \leq 3 \| |\phi _1\rangle \langle \phi _1| - |\phi _2\rangle \langle \phi _2| \|_2.\nonumber
\end{equation}

Finally, the Frobenius norm needs to be related to the Euclidean distance between $|\phi_1\rangle$ and $|\phi_2\rangle$. Note that

\begin{equation}
\| |\phi _1\rangle \langle \phi _1| - |\phi _2\rangle \langle \phi _2|\|_2 = \sqrt{2}\sqrt{1-|\langle \phi_1 |\phi_2\rangle|^2}\nonumber
\end{equation}

\noindent and,

\begin{equation}
\| |\phi _1\rangle - |\phi _2\rangle\|_2 = \sqrt{2}\sqrt{1-\text{Re}\left(\langle \phi_1 |\phi_2\rangle\right)}.\nonumber
\end{equation}

\noindent Hence,

\begin{equation}
\| |\phi _1\rangle \langle \phi _1| - |\phi _2\rangle \langle \phi _2|\|_2 \leq  \sqrt{2}\sqrt{1-\text{Re}\left(\langle\phi_1 | \phi_2\rangle\right)}\sqrt{1+\text{Re}\left(\langle\phi_1 | \phi_2\rangle\right)} \leq \sqrt{2}\| |\phi _1\rangle - |\phi _2\rangle\|_2.\nonumber
\end{equation}

\noindent Therefore,

\begin{equation}
|\gatefill(|\phi_1\rangle) - \gatefill(|\phi_2\rangle)| \leq 3\sqrt{2} \| |\phi _1\rangle - |\phi _2\rangle\|_2, \label{eq:Lipschitz}
\end{equation}

\noindent and so $3\sqrt{2}$ is a Lipschitz constant for $\gatefill : (\mathbb{CP}^{d-1},\| \: \|_2) \rightarrow \mathbb{R}$ which proves the theorem.

\end{proof}

\end{theorem}

\bigskip

For d fixed, the infimum over all such K is called the Lipschitz seminorm of $\gatefill$ and is denoted by $\eta$. An obvious corollary of the above theorem is that $\eta$ is bounded above by $3\sqrt{2}$. 
The metric space isomorphism between $(\mathbb{S}^{2d-1}, \| \: \|_2)$ and the set of unit vectors in $\mathbb{C}^d$ gives the following corollary,

\bigskip

\begin{corollary}

The function $\gatefill : (\mathbb{S}^{2d-1},\|\:\|_2) \rightarrow [0,1]$ is $3\sqrt{2}$-Lipschitz.

\end{corollary}

\bigskip

As discussed in section \ref{sec:BackConcentration} this implies that for $\epsilon > 0$,

\begin{equation}
\mathbb{P}_{\mu_F} \left[\gatefill \in \left(\overline{\gatefill}-\epsilon, \overline{\gatefill}+\epsilon \right)\right] \geq 1-4e^{\frac{-d\epsilon ^2}{81\pi ^3 ln2}}.\label{eq:measbound}
\end{equation}

\noindent Hence, if $\epsilon > 0$, and $|\phi\rangle$ is chosen randomly from the FS measure, the probability that the fidelity between $\Lambda (|\phi\rangle\langle \phi |)$ and $|\phi\rangle\langle \phi |$ is not $\epsilon$-close to the average is exponentially small in d, ie.

\begin{equation}
\text{pr}\left[\text{tr}\left(\Lambda\left(\kett\braa\right)\kett\braa\right) \in \left(\overline{\mathcal{F}_{\Lambda,\mathcal{I}}}-\epsilon, \overline{\mathcal{F}_{\Lambda,\mathcal{I}}} + \epsilon\right)\right] \geq 1-4e^{\frac{-d\epsilon ^2}{81\pi ^3 ln2}}.\label{eq:probbound}
\end{equation}

\subsection{Estimates and Bounds for the Average and Variance of the Gate Fidelity}\label{sec:AveandVariance}

The results of the previous section imply that the number of trials required to estimate the average gate fidelity between an unknown quantum operation $\Lambda$ and $\mathcal{I}$ decreases significantly as d grows large. 
Unfortunately generating Haar-random pure states is an inefficient task. It would therefore be useful to derive deviation inequalities similar to those given above for discrete sets of states with the counting measure. A natural set of states to analyze in this context are state k-designs~\cite{RBSC}, in particular approximate state 1 and 2-designs due to their ability to be efficiently generated~\cite{AE}.

A state k-design consists of states spread uniformly enough throughout $\pure$ so that the k'th central moment of the gate fidelity over the t-design is equal to the k'th central moment over $\pure$. An approximate state k-design is a finite set of states that approximates the k'th central moment over $\pure$ well. From equation (\ref{eq:probbound}) one would expect that in large dimensions, choosing a state uniformly at random from an approximate k-design would provide a good estimate of the average fidelity with high probability.

As mentioned previously, an explicit upper bound on $\sigma^2(\Lambda)$ is given by equation (\ref{eq:varupperbound}) which shows that $\sigma^2(\Lambda)$ scales as $O\left(\frac{1}{d}\right)$. One can also use the concentration results derived above to deduce both the asymptotic order of $O\left(\frac{1}{d}\right)$ for $\sigma^2(\Lambda)$ as well as an explicit upper bound that holds for every d. The method has the advantage of not requiring an exact expression for the variance and therefore is much simpler to obtain. The downside is that the upper bound is not as tight. For ease of notation, $\sigma^2(\Lambda)$ will be denoted by $\sigma^2$ throughout the rest of the presentation.

The asymptotic order of $\sigma^2$ 
is obtained by using (\ref{eq:measbound}) and Chebyshev's inequality which states that for any $k>0$,

\begin{equation}
\mathbb{P}_{\mu_F} \left[\gatefill \in \left(\overline{\gatefill}-k\sigma, \overline{\gatefill}+ k\sigma \right)\right] \geq 1-\frac{1}{k^2}.\nonumber
\end{equation}

\noindent From (\ref{eq:measbound}), any $\sigma > 0$ that satisfies the above equation for all d and $k>0$ must scale as $O\left(\frac{1}{\sqrt{d}}\right)$. Therefore the variance $\sigma^2$ scales as $O\left(\frac{1}{d}\right)$.

For $\epsilon > 0$ let $A_{\epsilon}$ denote the set $\gatefill \in \left(\overline{\gatefill}-\epsilon, \overline{\gatefill}+\epsilon \right)$. An upper bound on $\sigma^2$ can be found by noting that for any $\epsilon > 0$,

\begin{equation}
\sigma^2 = \mathbb{E}_{\mu_F}\left[\left(\gatefill-\overline{\gatefill}\right)^2 \: \Id_{A_{\epsilon}}\right] + \mathbb{E}_{\mu_F}\left[\left(\gatefill-\overline{\gatefill}\right)^2 \: \Id_{\pure/{A_{\epsilon}}}\right] \leq \epsilon^2 + 4e^{\frac{-d\epsilon ^2}{81\pi ^3 \ln (2)}},\nonumber
\end{equation}


\noindent where $\Id_A$ is the indicator function on $\pure$ with support on A, and similarly for $\Id_{\pure/A}$. Minimizing with respect to $\epsilon$ and defining $C=\frac{1}{81\pi^3 \ln(2)}$ gives,

\begin{equation}
\epsilon=\sqrt{\frac{\text{ln}(Cd)}{Cd}}.\nonumber
\end{equation}

\noindent Hence

\begin{gather}
\sigma^2\leq \frac{4+\text{ln}(Cd)}{Cd}\nonumber
\end{gather}

\noindent and so for n qubits,

\begin{equation}
\sigma^2 \leq \frac{4+\text{ln}(C)+\frac{n}{\text{ln}(2)}}{C2^{n}}.\nonumber
\end{equation}

\noindent As an example, for a 50 qubit system the above gives $\sigma^2 \leq 1.1 \times 10^{-10}$.  On the other hand equation (\ref{eq:varupperbound}) gives a tighter bound of $1.0 \times 10^{-14}$. Clearly for systems capable of performing large-scale quantum computations the variance of the gate fidelity will be extremely small.

\subsection{Convergence to Depolarization}\label{sec:Convergence}

This section will bring together many of the results from the previous sections as a single result: the asymptotic convergence to depolarization of quantum channels with respect to the gate fidelity. The convergence is quantified in two ways, the first utilizing the $L^2$ metric on the set $\xi$ of gate fidelity random variables and the second resembling the notion of convergence in probability.

If $\mathcal{G}$ and $\mathcal{K}$ are two quantum operations on $\Linops$ then the $L^2$ distance, denoted here by $d_2$, between $\mathcal{F}_{\mathcal{G},\mathcal{I}}$ and $\mathcal{F}_{\mathcal{K},\mathcal{I}}$ is,

\begin{equation}
d_2\left(\mathcal{F}_{\mathcal{G},\mathcal{I}},\mathcal{F}_{\mathcal{K},\mathcal{I}}\right)=\left(\mathbb{E}_{\mu_F}\left[\left(\mathcal{F}_{\mathcal{G},\mathcal{I}}-\mathcal{F}_{\mathcal{K},\mathcal{I}}\right)^2\right]\right)^{\frac{1}{2}}.\nonumber
\end{equation}

\noindent Suppose that $\mathcal{G}$ has average fidelity equal to $b$ and that $\mathcal{K}$ is the depolarizing channel with (constant) gate fidelity equal to $b$. Denoting $\mathcal{K}$ by $\mathcal{G}_{\text{dep}}$,

\begin{equation}
d_2\left(\mathcal{F}_{\mathcal{G},\mathcal{I}},\mathcal{F}_{\mathcal{G}_{\text{dep}},\mathcal{I}}\right)=\left(\mathbb{E}_{\mu_F}\left[\left(\mathcal{F}_{\mathcal{G},\mathcal{I}}-b\right)^2\right]\right)^{\frac{1}{2}}\nonumber
\end{equation}

\noindent which is just the standard deviation of $\mathcal{F}_{\mathcal{G},\mathcal{I}}$. Therefore from equation (\ref{eq:varupperbound}), for every $d$,

\begin{equation}
d_2\left(\mathcal{F}_{\mathcal{G},\mathcal{I}},\mathcal{F}_{\mathcal{G}_{\text{dep}},\mathcal{I}}\right) \leq \sqrt{\frac{8d^3 + 16d^2 + 4d}{\left(d^2+2d+1\right)\left(d^2+5d+1\right)}}.\nonumber
\end{equation}

\noindent and so $d_2\left(\mathcal{F}_{\mathcal{G},\mathcal{I}},\mathcal{F}_{\mathcal{G}_{\text{dep}},\mathcal{I}}\right) \rightarrow 0$ as $O\left(\frac{1}{\sqrt{d}}\right)$.

The second method uses the concentration of measure results from section \ref{sec:Concentration}. It is straightforward to turn equation (\ref{eq:measbound}) into a statement regarding convergence to depolarization by noting that since $\mathcal{F}_{\mathcal{G}_{\text{dep}},\mathcal{I}}$ is constant and equal to $b$, for any $\epsilon > 0$,

\begin{equation}
\mathbb{P}\left[\left|\mathcal{F}_{\mathcal{G},\mathcal{I}}-\mathcal{F}_{\mathcal{G}_{\text{dep}},\mathcal{I}}\right| \leq \epsilon\right] \geq 1-4e^{\frac{-d\epsilon ^2}{81\pi ^3 \ln(2)}}.\nonumber
\end{equation}

\noindent Hence for $\epsilon > 0$ fixed,

\begin{equation}
\lim_{d\rightarrow \infty}\mathbb{P}\left[\left|\mathcal{F}_{\mathcal{G},\mathcal{I}}-\mathcal{F}_{\mathcal{G}_{\text{dep}},\mathcal{I}}\right| \leq \epsilon\right] = 1.\nonumber 
\end{equation}





\subsection{Estimating the Minimum Gate Fidelity}\label{sec:Minimum}

In this section methods are discussed for estimating the minimum of the gate fidelity. The first method uses the Lipschitz constant given by equation (\ref{eq:Lipschitz}) and the existence of fine ``nets" on the set of pure states. The second method uses the bound given by equation (\ref{eq:measbound}).

Nets of states are defined as follows: If $\epsilon > 0$ and $g$ is a metric on $\mathbb{CP}^{d-1}$, an $(\epsilon,g)$-net is defined to be a finite set of states $\mathcal{N}_{(\epsilon,g)} \subset \mathbb{CP}^{d-1}$ such that for any $|\psi\rangle \in \mathbb{CP}^{d-1}$ there exists $|\phi\rangle \in \mathcal{N}_{(\epsilon,g)}$ satisfying

\begin{equation}
g\left(|\psi\rangle,|\phi\rangle\right)\leq \epsilon.\nonumber
\end{equation}





\noindent It has been shown~\cite{MS} that for $\epsilon \in (0,1)$ and $g$ induced by the 1-norm there exists an $(\epsilon,\| \: \|_1)$-net such that

\begin{equation}
\left|\mathcal{N}_{(\epsilon,\| \: \|_1)}\right| \leq \left(\frac{5}{\epsilon}\right)^{2d}.\label{eq:netbound}
\end{equation}

\noindent This particular net is also shown to be a $(\frac{\epsilon}{2},\| \: \|_2)$-net.

Let $\epsilon >0$ and put $\| \: \|_2$ on $\mathbb{CP}^{d-1}$. From above, there exists an $\mathcal{N}_{(\epsilon,\| \: \|_2)}$ net of size $\left(\frac{5}{2\epsilon}\right)^{2d}$ on $\pure$. Suppose the minimum of the gate fidelity over $\mathbb{CP}^{d-1}$ occurs at $|\psi\rangle$. By definition there exists a state $|\phi\rangle \in \mathcal{N}_{(\epsilon,\| \: \|_2)}$ such that

\begin{equation}
\||\psi\rangle - |\phi\rangle \|_2 \leq \epsilon.\nonumber
\end{equation}

\noindent Using the Lipschitz condition in equation (\ref{eq:Lipschitz}), if $\Lambda$ is a quantum operation,

\begin{equation}
\left|\gatefill\left(|\psi\rangle\right)-\gatefill\left(|\phi\rangle\right)\right| \leq 3\sqrt{2}\||\psi\rangle - |\phi\rangle \|_2,\nonumber
\end{equation}

\noindent which implies



\begin{equation}
\gatefill\left(|\phi\rangle\right) - 3\sqrt{2}\epsilon \leq \gatefill\left(|\psi\rangle\right).\label{eq:minest}
\end{equation}

\noindent Therefore the minimum of $\gatefill$ over $\mathbb{CP}^{d-1}$ is bounded below by $\gatefill\left(|\phi\rangle\right) - 3\sqrt{2}\epsilon$ and the minimum over the net is a good approximation to the minimum over the entire space.

As mentioned previously, by a simple concavity argument, the minimum of the gate fidelity over all mixed input states occurs at a pure state. Therefore equation (\ref{eq:minest}) provides an estimate for the minimum over all mixed states. With the bound on the size of $\mathcal{N}_{(\epsilon,\| \: \|_2)}$ given in equation (\ref{eq:netbound}), this method will only be useful for small quantum systems. More scalable bounds on the size of the net would imply the applicability of this method for larger quantum systems.



Property 2 from~\cite{GLN} (see section \ref{sec:gatefilanddist}) is that a useful distance measure should be easy to calculate. The minimum gate fidelity has the drawback of not being easy to calculate analytically, even when a description of the noise process is available. However, convex optimization techniques can be used to numerically evaluate an estimate for the minimum when the noise process is known. The above lower bound implies that if one has a description of the noise then evaluating the minimum fidelity over a finite set of states gives an approximation of the minimum over all mixed quantum states. Tightening the bounds on the size of the net required would make this method more applicable.

This method also gives a clear experimental procedure for estimating the minimum gate fidelity (property 3 from~\cite{GLN}) without requiring process tomography. The idea is to be able to prepare a suitable net of states and determine the minimum fidelity over these states by performing measurements in the appropriate bases. Again, this minimum provides a good approximation to the minimum over all states but the obvious drawback is that the number of states scales poorly with the dimension of the system.

The second method for estimating the minimum gate fidelity uses the concentration result for the gate fidelity given in equation (\ref{eq:measbound}). 
 Let $Q>0$ be fixed and suppose one is only interested in finding the smallest value $\gatefill$ can take such that any state $|\phi\rangle$ producing a smaller value lies in a set whose measure equals $Q$. In this context the smallest value is called the effective minimum, denoted $\mathcal{F}_{\text{eff}},$ given the tolerance $Q$. This problem is equivalent to finding the maximum over all b $\in [0,1]$ satisfying,

\begin{equation}
\mathbb{P}_{\mu_F}\left[\gatefill \in \left[0,b\right]\right] \leq Q.\nonumber
\end{equation}

\noindent The maximum value of b is equal to $\mathcal{F}_{\text{eff}}$ and depends on both $d$ and $Q$.

By equation (\ref{eq:measbound}) for every $\epsilon > 0$,

\begin{equation}
\mathbb{P}_{\mu_F}\left[\gatefill \in \left[0,\overline{\gatefill}-\epsilon\right]\right] \leq 2\text{exp}\left(\frac{-d\epsilon^2}{81\pi^3\text{ln}(2)}\right).\nonumber
\end{equation}

\noindent This inequality can be used to find a non-trivial lower bound for b. Let $\epsilon_{Q,d}$ be the value of $\epsilon$ obtained when $Q=2\text{exp}\left(\frac{-d\epsilon^2}{81\pi^3\text{ln}(2)}\right)$,

\begin{equation}
\epsilon_{Q,d}=\sqrt{\frac{81\pi^3\text{ln}(2)\: \text{ln}\left(\frac{2}{Q}\right)}{d}}.\nonumber
\end{equation}

\noindent By construction $\epsilon_{Q,d}$ satisfies $\mathbb{P}_{\mu_F}\left[\gatefill \in \left[0,\overline{\gatefill}-\epsilon_{Q,d}\right]\right] \leq Q$ and so by definition,

\begin{equation}
\overline{\gatefill} \geq \mathcal{F}_{\text{eff}} \geq \overline{\gatefill}-\epsilon_{Q,d} = \overline{\gatefill}-\sqrt{\frac{81\pi^3\text{ln}(2)\: \text{ln}\left(\frac{2}{Q}\right)}{d}}. \nonumber
\end{equation}

\noindent This lower bound on $\mathcal{F}_{\text{eff}}$ is non-trivial since for fixed $Q$, $\epsilon_{Q,d} \rightarrow 0$ as $d \rightarrow \infty$. Therefore $\mathcal{F}_{\text{eff}} \rightarrow \overline{\gatefill}$ as $d\rightarrow \infty$, and the effective minimum and average of the gate fidelity become indistinguishable for large $d$.





\section{Conclusion and Further Research}\label{sec:Conclusion}

If $\mathcal{Q}$ is a full-rank quantum operation and $d\geq 4$ it has been shown that there exist quantum channels $\mathcal{R}$ not equal to $\mathcal{Q}^{\dagger}$ which produce the same gate fidelity function as $\mathcal{Q}$. A corollary of this result is that when $d \geq 4$ and $\mathcal{Q}$ is a depolarizing channel, there exist non-depolarizing channels $\mathcal{R}$ which produces the \textit{same} gate fidelity as $\mathcal{Q}$. Since $\mathcal{Q}$ has a constant gate fidelity on $\pure$, there exist non-depolarizing channels with a constant gate fidelity on $\pure$.

Intuitively, the fact that theorem \ref{thm:nonuniqueness} holds in higher dimensions seems to be related to the rich geometry of the Bloch space representation of quantum states in
higher dimensions. The simple Bloch sphere representation of a single qubit appears to indicate that theorem \ref{thm:nonuniqueness} cannot be extended to $d=2$ however this, along with the
status of $d=3$, remain open questions. 
An entire family of open questions arising from theorem \ref{thm:nonuniqueness} relates to how two quantum channels which produce the same gate fidelity can differ with respect to a specific information-theoretic property. For instance, an interesting direction of research would be to analyze whether two quantum channels which produce the same gate fidelity can differ in their capacities for transmitting information.

Using Levy's lemma, an upper bound on the probability that a randomly chosen pure state produces a gate fidelity value far from the average has been derived. The upper bound converges to 0 exponentially quickly in the number of qubits comprising the quantum system. Hence in large dimensions very few trials are required to estimate the average of the gate fidelity to high accuracy. Extending the result to approximate state k-designs would be useful due to their ability to be efficiently generated~\cite{AE}. An upper bound on the variance of the gate fidelity is obtained which implies that all quantum channels converge to depolarizing channels with respect to the gate fidelity as $d \rightarrow \infty$.

Two methods for estimating the minimum of the gate fidelity have been presented, one using the Lipschitz condition on the gate fidelity function, and the other using the concentration inequalities obtained from Levy's lemma. The first method shows that the minimum over a suitably large net of pure states will be a good approximation to the minimum over all mixed quantum states. Improvements on the size of the net would make the method more applicable in larger dimensions. The second method gives estimates for the minimum up to a tolerated measure of deviating states. As expected from the deviation inequalities, this effective minimum becomes indistinguishable from the average as the dimension of the system grows large.

\nonumsection{Acknowledgements}
\noindent
The author would like to thank John Watrous, Marco Piani, Joseph Emerson, David Kribs and Yingkai Ouyang for helpful discussions and acknowledges financial support from NSERC and CIFAR.

\end{document}